# Modelling attenuation and velocity of ultrasonics in reconstituted milk powder

Bernard Lacaze
TéSA, 9 Bd de la Gare, Toulouse, France
*e-mail*: bernard.lacaze@tesa.prd.fr

May 6, 2020

**Abstract**

In the context of food quality control, ultrasonics provide proven methods which are able to replace manual controls. The latter are subject to the lack of objectivity of human judgement. Automatic control increases reliability and reduces costs.

This paper revisits data coming from ultrasonics through reconstituted milk powder. Two characteristics are studied using five productions of a well known manufacturer. Measured attenuation and dispersion of ultrasonics are explained through stable probability laws and random propagation times. We have proved elsewhere that this model is available in many propagation problems, in acoustics, ultrasonics and in the electromagnetic world.

*keywords:* ultrasonics, instant milk powder, linear filtering, stable probability law, random propagation times.

## 1 Introduction

### 1.1 Instant milk powder visual reconstruction test

When milk powder is mixed with water, the result is a liquid which homogeneity is blurred by liquid-fat aggregates at the surface and small particles of protein gel in the bulk. To evaluate the impact of these flaws on consumers, visual reconstitution tests (RT) are carried out. People give a rating to some number of samples coming from different sources. In [1] (forming the basis of this article), five samples were examined, from five different milk powders, by an undetermined number of people. Each of them received two notes between 0 and 5, one for the surface appearance (ST) and the other for the bulk appearance (BT). The table below gives the test results.

|    | 1   | 2   | 3   | 4   | 5   |
|----|-----|-----|-----|-----|-----|
| $BT$ | 3.5 | 3   | 2.5 | 2   | 1   |
| $ST$ | 4   | 4.5 | 2.5 | 1.5 | 1.5 |

Table 1: visual reconstitution test

The numbering of samples was done a posteriori, from the worst (powders 1 and 2 are not acceptable) to the best ( 4 and 5 are acceptable) and 3 is questionable.

It is clear that BT and ST scores are linked: a high BT goes with a high ST and likewise for middle and low scores. This is corroborated by the estimation of the correlation coefficient equal to 0.9 (but limited to 5 samples). We know that its maximum value is 1, which implies a linear relation between both rows of table 1. Noneless, people agreed with the opinion that this kind of test is not very reliable, too dependent of the mood of examiners, and due to "*the lack of objectivity of the visual inspection step*". This leads to replacing people by machines and we can expect that methods based only on physical and/or chemical properties are better, at least stable and reliable (and less expensive).

Among them, ultrasonics provide methods of quality control in many manufacturing processes, and particularly in food production [2], [3]. In particular, paper [1] studies the propagation of ultrasonics through media providing both visual tests ST and BT.

## 1.2 Milk powder and ultrasonics

Measurements of attenuation and velocity of ultrasonics in the frequency band 3-7MHz are reported in [1]. Here, figures 1a, 1b, 2a, 2b utilize figures 3 to 6 of [1]. The first two figures 1a, 1b, are linear regressions about velocities, but their reliability is questionable due to non-negligible curvatures (for instance for samples 2, 3, 4). Figures 2a, 2b express ln$att$ as function of ln$f$ ($f$ in MHz), in accordance with the usual attenuation in $f^\alpha$.

Each figure contains 5 curves, and data of [1] show that it is reasonable to approximate with straight lines (in spite of doubts about velocity $v(f)$). Authors assert that "*these experiments were repeated with another set of powder from another factory and similar results were obtained*". Relative positions of lines in figures 2a and 2b are in accordance with the order in table 1. This is incitative to replacing visual tests by measures of attenuation of ultrasonics. On the contrary, lines of velocities $v(f)$ in figures 1a, 1b intersect, which prohibits the same procedure.

In this paper, we propose a model of propagation where $v^{-1}(f)$ is a linear function of $f^{\alpha-1}$ or ln$f$ (when $\alpha = 1$). $\alpha$ is the parameter defined below in (1) and it characterizes the attenuation shape. Experiments in [1] are in accordance with this model (see figures 3a, 3b, 3c, 3d). Equivalently, the complex gain of the linear invariant filter $\mathcal{F}$ which summarizes the propagation is the characteristic function of a stable probability law (see sections 2.3 and 3.1). The fact that one of parameters is close to $\pm 1$ in all cases, shows that the causality property is well approached (section 3.2).

Finally, we explain why random propagation times following stable probability laws provides a good framework for low power ultrasonics used in food industry and particularly dairy products.

# 2 Modelling the ultrasonic propagation

## 2.1 Attenuation

In ultrasonics, it is wellknown that attenuations are most of the time in the form (see [4] to [9])
$$att\,[f] = c\,(2\pi f)^\alpha \qquad (1)$$

$\alpha$ and $c$ are characteristics of the crossed medium. $\alpha = 2$ for water and atmosphere [10], [11] (it is the highest value) but $\alpha$ can take any value between 0 and 2. Figures 2a and 2b deduced from figures 5 and 6 of [1] illustrate this property for milk powder under the linear form (with respect to $\ln f$)

$$\ln att\,[f] = \alpha \ln f + \ln c + \alpha \ln(2\pi) \tag{2}$$

despite gaps close to 7MHz. Table 2 gives estimations of $\ln att\,[f]$ for each sample BT and ST

| $\ln att\,[f]$ | BT | ST |
|---|---|---|
| 1 | $0.86 \ln f + 3.24$ | $1.13 \ln f + 2.51$ |
| 2 | $0.87 \ln f + 3.31$ | $\ln f + 2.6$ |
| 3 | $\ln f + 3.27$ | $1.1 \ln f + 2.73$ |
| 4 | $0.85 \ln f + 3.45$ | $0.85 \ln f + 3.5$ |
| 5 | $0.77 \ln f + 3.59$ | $0.87 \ln f + 3.48$ |

Table 2: equations of lines in figures 2a and 2b about attenuations.

Equality (1) is an empirical relation which is simpler than an approximation by a polynomial of degree larger than 1. Physical justifications are not given for approximation (1), except in rare cases, for instance $\alpha = 2$ for atmosphere and water, or $\alpha = 1/2$ for high frequency propagation through electrical cables [12], [13]. Lines of figures 2a and 2b have slopes ($\alpha$ values) around 1 and different enough values of $c$ which lead to separate curves without intersection. Their respective positions are in accordance with visual tests.

The concordancy between the relative place of curves in figures 2a, 2b, and the rank in visual tests cannot be due to chance. The probability of such a coincidence is equal to 1/120 for 5 products (like here), 1/24 for 4 or 1/720 for 6. Consequently, replacing visual tests by measure of attenuation is not a very risky gamble.

## 2.2 Velocity

Figures 1a and 1b are obtained from data in figures 3 and 4 of [1]. They are linear regressions giving velocity $v(f)$ (in $m.s^{-1}$) as a function of frequency $f$ (in MHz). According to the authors, the velocity dispersion is measured with an accuracy of 0.1-0.2 $m.s^{-1}$ which seems optimistic to me. Linear regressions do not take into account curvatures in data of [1], which can be hidden by inaccuracies of measurements. Moreover, the propagation time on the unit distance is $v^{-1}(f)$. It is a quantity as interesting as the celerity $v(f)$, which justifies the research of regressions of $v^{-1}(f)$.

For reasons explained below, it is of interest to highlight a linear dependence between $v^{-1}(f)$ and $f^{\alpha-1}$ or $\ln f$ (when $\alpha = 1$), where $\alpha$ is the slope in (2). We will accept the following equalities

$$v^{-1}(f) = \begin{cases} \gamma f^{\alpha-1} + m, \alpha \neq 1 \\ \gamma \ln f + m, \alpha = 1 \end{cases} \tag{3}$$

which link attenuation and dispersion through $\alpha$. Figures 3a, 3b, 3c, 3d prove this property for data in [1]. Fig3a corresponds to BT1 ($\alpha = 1.14$), Fig3b to BT3 ($\alpha = 1$), Fig3c to ST1 ($\alpha = 1.13$) and Fig3d to ST5 ($\alpha = 0.87$). Table 3 contains the whole set of cases ($f$ in MHz and results in $\mu$s.m$^{-1}$).

| $v^{-1}(f)$ | BT | ST |
|---|---|---|
| 1 | $16.7f^{-0.14} + 405$ | $-10.8f^{0.13} + 435$ |
| 2 | $21.3f^{-0.13} + 401$ | $-1.71\ln f + 421$ |
| 3 | $-2.2\ln f + 427$ | $-17.5f^{0.1} + 431$ |
| 4 | $19.6f^{-0.15} + 403$ | $13.9f^{-0.15} + 407$ |
| 5 | $11.4f^{-0.23} + 412$ | $21.8f^{-0.13} + 401$ |

Table 3: estimations of parameters
$\gamma$ and $m$ of equality (3)

Formulas (3) are linked to causality (Kramers-Kronig relations)[7], [14], [15], and to stable probability laws [13], [16], [17], see section 3. Tables 4 and 5 summarize both linear forms for BT and ST.

| BT | $\alpha$ | $c$ | $\gamma$ | $m$ |
|---|---|---|---|---|
| 1 | 0.86 | 5.3 | 16.7 | 405 |
| 2 | 0.87 | 5.1 | 21.3 | 401 |
| 3 | 1 | 4.1 | $-2.2$ | 427 |
| 4 | 0.85 | 6.6 | 19.6 | 403 |
| 5 | 0.77 | 8.6 | 11.4 | 412 |

Table 4: parameters values
for bulk samples, $m$ in $\mu$s.m$^{-1}$

| ST | $\alpha$ | $c$ | $\gamma$ | $m$ |
|---|---|---|---|---|
| 1 | 1.13 | 1.57 | $-10.8$ | 435 |
| 2 | 1 | 2.12 | $-1.71$ | 421 |
| 3 | 1.1 | 2.05 | $-17.5$ | 430 |
| 4 | 0.85 | 6.05 | 13.9 | 407 |
| 5 | 0.87 | 5.64 | 21.8 | 401 |

Table 5: parameters values for
surface samples, $m$ in $\mu$s.m$^{-1}$

Tables 4 and 5 give no obvious information about the quality of samples (except perhaps the parameter $c$ for ST), though they summarize data of [1]. In the case of ST samples, values of $\alpha, c$, are very different for ST1, ST2, with respect to ST4 and ST5, which could justify a split in two groups. But values for ST3 are questionable about a membership to the first group.

## 2.3 Equivalent linear invariant filter

Let $g(t) = \exp[2i\pi ft]$ be the monochromatic wave at the frequency $f > 0$. The crossing of reconstituted milk provides the same kind of wave with some amplitude and some phase depending on the frequency $f$. It is equivalent to consider a Linear Invariant Filter (LIF) $\mathcal{F}$ such as (for a unit thickness)

$$\mathcal{F}[g](t) = \exp\left[-att[f] + 2i\pi f\left(t - \frac{1}{v(f)}\right)\right] \qquad (4)$$

where $v^{-1}(f)$ is the propagation time in seconds and $attf$ is given in Neper (for a distance of one meter). This means that the result is the monochromatic wave at the frequency $f$ delayed by $v^{-1}(f)$ and weakened by $\exp[-att(f)]$. Therefore, $-att(f)$ is the neperian logarithm of the weakening (used in [1]).

Using $(2)$ and $(3)$, the complex gain (or frequency response) $F$ of $\mathcal{F}$ is [18], [19]

$$F(f) = \begin{cases} \exp\left[-c(2\pi f)^\alpha - 2i\pi f\left(m + \gamma f^{\alpha-1}\right)\right], \alpha \neq 1 \\ \exp\left[-c(2\pi f)^\alpha - 2i\pi f\left(m + \gamma \ln f\right)\right], \alpha = 1. \end{cases} \quad (5)$$

If $g(t) = \exp[2i\pi ft]$ is the input of the LIF $\mathcal{F}$, $\mathcal{F}[g](t) = F(f)e^{2i\pi ft}$ is the output of $\mathcal{F}$. In the case of complex gains as $(5)$, we will explain (see section 3) why the most favourable situations correspond to parameters linked as

$$\gamma \cong c(2\pi)^{\alpha-1} \tan\frac{\pi\alpha}{2} \text{ for } \alpha \neq 1 \text{ and } \gamma \cong -\frac{2c}{\pi} \text{ for } \alpha = 1.$$

To summarize, the propagation of ultrasonics through diluted milk powder can be modelled by a LIF $\mathcal{F}$ of complex gain $F(f)$ defined by (5) from parameters $(\alpha, c, \gamma, m)$ which characterize the attenuation $att[f]$ and the propagation time $v^{-1}(f)$. In section 3 and 4 below, we explain links with stable probability laws and with random propagation times.

## 3 Stable probability laws

### 3.1 Definition

Let assume that $A_1, A_2...$ are independent (real) random variables (r.v) following some probability law $\mathcal{L}$. It is a stable probability law if linear combinations of the $A_n$ follows the same law $\mathcal{L}$ (excluding location and scale parameters). The Gaussian law is stable, and it is the only one with two finite moments (the mean and the variance). The real r.v $A$ follows a stable law of parameters $(\alpha, c, \beta, m')$ if its characteristic function has the shape $(\omega > 0)$

$$\mathrm{E}\left[e^{-i\omega A}\right] = \exp\left[-im'\omega - c|\omega|^\alpha(1 + i\beta\theta(\omega))\right]$$

$$\theta(\omega) = \begin{cases} \tan(\pi\alpha/2) \text{ if } \alpha \neq 1 \\ (2/\pi)\ln\omega \text{ if } \alpha = 1 \end{cases} \quad (6)$$

with $0 < \alpha \leq 2, -1 \leq \beta \leq 1, c > 0$, real $m'$, and a continuation for $\omega < 0$, using the Hermitian symmetry [16], [20], [21]. Stable laws have probability densities, and are unimodal. The probability density $\mu_{m'}(x)$ corresponding to the characteristic function $\mathrm{E}\left[e^{-i\omega A}\right]$ verifies ([16] th.3.2.2)

$$\mu_{m'}(x) = \frac{1}{2\pi}\int_{-\infty}^{\infty} \mathrm{E}\left[e^{-i\omega A}\right] e^{i\omega x} d\omega. \quad (7)$$

Clearly, (5) and (7) are confused when $(\omega = 2\pi f)$,

$$\begin{cases} c\beta(2\pi)^{\alpha-1}\tan(\pi\alpha/2) = \gamma, m' = m \text{ if } \alpha \neq 1 \\ \frac{2}{\pi}c\beta = \gamma, m' + \frac{2}{\pi}c\beta\ln(2\pi) = m \text{ if } \alpha = 1. \end{cases} \quad (8)$$

Consequently, effects of ultrasonics on milk powder can be explained in the context of stable probability laws provided that parameters stay within acceptable limits. It is the case for $\alpha$ ($0.8 < \alpha < 1.2$ see tables 4 and 5), and for $c$ ($c > 0$). Table 6 gives values of $\beta$ computed from measured values of $\alpha, c, \gamma$ (see tables 4 and 5). For BT (bulk samples), we find $\alpha \leq 1, |\beta| > 0.7$ with one value at 1.08 which may be lowered to 1. For ST (surface samples), values of $\beta$ are larger (for the first three), up to exceeding the limit $|\beta| = 1$.

| $\beta$ | 1 | 2 | 3 | 4 | 5 |
|---|---|---|---|---|---|
| $BT$ | 0.91 | 1.08 | $-0.86$ | 0.96 | 0.76 |
| $ST$ | 1.12 | $-1.26$ | 1.12 | 0.73 | 1.02 |

Table 6: values of $\beta$ from data in [1]

Estimation errors of parameters $\alpha, c, \gamma, m$ can explain anomalous values of $\beta$.

## 3.2 Causality

A stable law is symmetric (with respect to $m'$) when $\beta = 0$. On the other hand, the stable law is one-sided only when $\alpha < 1, \beta = \pm 1$. For $\alpha < 1, \beta = 1$ we have $\mu_{m'}(x) = 0$ for $x < m'$ [16]. Elsewhere, $\beta$ is the parameter which rules asymmetry. The larger is $|\beta|$, the larger is the asymmetry of the probability density $\mu_{m'}(x)$. For $\alpha > 1$ (resp. $\alpha = 1$), the causality is approached when $\beta$ is close to 1 (resp. close to -1). When $\beta > 1$, the impulse response $\mu_{m'}(x)$ loses the positive character of a probability density, but the near causality is not affected.

In practical situations, parameters are estimated, and the true value of $\beta$ is never reached (except by chance). So, we never find exactly $\beta = \pm 1$ and a strict causality. A positive value of $\beta$, when $\alpha \neq 1$, is the sign of a probability law with a fat tail on the positive axis, and weak probability on the negative axis. A positive $m'$ moves probability masses towards the right, and then lightens them on the negative axis. The addition of both properties leads to an approximate causality [8].

When we approximate the complex gain (5) with the stable law (7), relation (8) links the parameters $\gamma$ and $\beta$ through $\alpha$ and $c$. Data in [1] show that estimations of $\beta$ remain in a neighbourhood of $\pm 1$ (table 6). Equivalently, from (9), we always have $\gamma$ close to $c(2\pi)^{\alpha-1} \tan \frac{\pi\alpha}{2}$ for $\alpha \neq 1$ and $-2c/\pi$ for $\alpha = 1$. This means that the attenuation (governed by $\alpha, c$) approximately determines variations of the celerity (governed by $\alpha, \gamma$) and vice versa. This property is linked to the Kramers-Krönig relations [7], [14], [15].

Equivalently, from (8), the filter $\mathcal{F}$ defining the ultrasonic propagation is causal only when

$$\alpha < 1, c(2\pi)^{\alpha-1} \tan(\pi\alpha/2) = \gamma. \tag{9}$$

The strict causality is obtained at conditions which cannot be verified because the parameters are computed from experiments which contain errors, and then the value $\beta = \pm 1$ cannot be exactly obtained. Noneless, the value of $\mu_0(x)$ is close to 0 for $x < x_0$, where $-x_0$ is close to a few units. Here, $m$ is always in the order of 400 ($v^{-1}$ in $\mu$s.m$^{-1}$), which implies that $\mu_m(x) = \mu_0(x - m)$ is always negligible for $x < 0$. We consider that the causality condition is fulfilled [8].

Figures 4a, 4b, 4c, show probability densities $\mu_0$ of stable probability laws which are the Fourier transforms (7) of (6). They are matched to cases ST2 ($\alpha =$

$1, c = 2$), BT2 ($\alpha = 0.87, c = 5$), ST1 ($\alpha = 1.13, c = 1.5$), with $\beta = 0.8, 0.9, 1$. The Fourier transforms are also given for values $\beta = 1.1, 1.2$, but they are no longer probability densities (they take negative values). To take into account the value of $m$ or $m'$ (around 400) is equivalent to move curves towards the right by approximately 400 units. We see the influence of $c$, which is a scale parameter (the curves for ST1 are more distinct). When it increases, it spread and separates the different curves.

## 4  Random propagation times and stable laws

In this section, we explain why LIF in ultrasonics can often be replaced by random propagation times [8], [9], [17]. We consider the random process $\mathbf{U} = \{U(t), t \in \mathbb{R}\}$ defined by

$$U(t) = e^{i\omega_0(t - B(t))} \tag{10}$$

where $\mathbf{B}(t) = \{B(t), t \in \mathbb{R}\}$ is a real process such that both following characteristic functions (in the probability sense) do not depend on $t$

$$\begin{cases} \psi(\omega) = \mathrm{E}\left[e^{-i\omega B(t)}\right] \\ \phi(\tau, \omega) = \mathrm{E}\left[e^{-i\omega(B(t) - B(t-\tau))}\right]. \end{cases} \tag{11}$$

They define the probability laws of the random variables (r.v.) $B(t)$ and $B(t) - B(t-\tau)$. The independence with respect to $t$ of these laws implies a stationarity stronger than the usual second order stationarity.

$\mathbf{U}$ models a monochromatic wave of frequency $f_0 = \omega_0/2\pi$ which has crossed a medium on some distance. $\mathbf{B}$ models the time spent by the wave. Because $\mathbf{B}$ has a random character, $\mathbf{U}$ is no longer monochromatic. It is not difficult to prove the following formulas [8], [13], [22]

$$\begin{array}{c} U(t) = G(t) + V(t), \qquad G(t) = \psi(\omega_0) e^{i\omega_0 t} \\ 2\pi s_V(\omega) = \int_{-\infty}^{\infty} \left[\phi(\tau, \omega_0) - |\psi(\omega_0)|^2\right] e^{-i(\omega - \omega_0)\tau} d\tau \end{array} \tag{12}$$

$\mathbf{G}$ is deterministic. It is the monochromatic wave which has been attenuated and dephased by $\psi(\omega_0)$. $\mathbf{V}$ is a zero-mean random process, stationary (in the wide sense), and with the spectral density $s_V(\omega)$. The integral in (12) is well defined provided that $\phi(\tau, \omega_0) \to_{\tau \to \infty} |\psi(\omega_0)|^2$ fast enough (this means that the random variables $B(t)$ and $B(t-\tau)$ are almost independent for large $\tau$). In a signal theory context, $\mathbf{G}$ is the output of the linear invariant filter (LIF) with complex gain (or frequency response) $\psi(\omega)$. $\mathbf{G}$ is monochromatic, but accompanied by a process $\mathbf{V}$ with band spectrum, which can be considered as an additive noise. The total power of $\mathbf{U}$ is one, and then the transformation from the monochromatic wave $e^{i\omega_0 t}$ to $\mathbf{U}$ verifies the theorem of energy balance: the whole transmitted energy is retrieved through the addition "signal" $\mathbf{G}$ + "noise" $\mathbf{V}$.

A liquid or a gas is made of molecules which move indefinitely, the difference coming from the type of links between molecules. The power is transmitted by random shocks. The result has to contain a random part even when the power to be transmitted is deterministic. $\mathbf{V}$ represents this random part which is not taken into account in measurements, because of a power spectrum which is too spread out (with respect to the frequency window of devices). Properties which define the process are the time between molecular shocks, the length of trajectories, or the number of shocks by time unit. These events are so numerous that the result **V** is like a noise with very high frequencies. This is why devices, which have a limited frequency window fitted to **G**, cannot see **V**.

Let's assume that $B(t)$ follows a stable law $(\alpha, c, \beta, m')$ as $A$ (section 3.1). The probability law of $B(t)$ is $\mu_{m'}(x)$ defined by (6) and (7). From (12), **G** is the output of a LIF of complex gain $\psi(\omega)$. As explained in section 3, **G** is a correct model for the studied propagation, and also **U=G+V,** because **V** cannot be viewed by devices.

The same model of propagation can be used for other frequency bands and other media, for instance acoustics [23], or light propagation [24], or propagation through electrical cables [13]... In some cases, the **G** part disappears, the **V** part is observed, but it is possible that both parts **G** and **V** remain visible [25].

# 5 Conclusion

Ultrasonics provide methods of quality control in many manufacturing processes, and particularly in food production [2], [3]. The beam attenuation and/or its velocity are measured and they give information which may be used to monitor factories. A conclusion of paper [1] about reconstructed milk is:

".. *the measurement of the ultrasonic attenuation coefficient can be very well correlated to visual scores, for both the ST and RT*" and

"*Ultrasonic velocity cannot be correlated to a visual score...*"

Authors added:

"*These experiments were repeated with another set of powder from another factory and similar results were obtained*".

Both citations prove that ultrasonics in some frequency band provide good information from the attenuation, and that a plan with five products is sufficient. We have shown that they allow to characterize the linear filter which models the propagation.

Table 1 of [1] is the result of visual examinations of diluted milk powders. Powders are ranked from the worst to the best, and results are more or less identical for bulk and surface appearances. Ultrasonics in the 3-7MHz band provide values of attenuations and velocities. Graphs of ln*att* in function of ln$f$ are in accordance with visual examinations (see figures 2a, 2b), and define regression lines of slope $\alpha$ with $0.7 < \alpha < 1.2$ (see table 2) and of parameter $c$ in accordance with the usual relation $att(f) = c(2\pi f)^{\alpha}$.

It is not the case for velocities (see figures 1a, 1b), perhaps due to the choice of coordinates $(f, v(f))$ to perform regression lines. The system of coordinates $(f^{\alpha-1}, v^{-1}(f)$ or $(\ln f, v^{-1}(f))$ when $\alpha = 1$, is the right choice for regression lines (see (3) and figures 3a to 3d). Parameters of both systems are closely linked through the values of $\alpha, c$ and $\beta$ (which is close to $\pm 1$).

The previous considerations allow to characterize the propagation by an equivalent linear invariant filter $\mathcal{F}$ and a system of parameters $(\alpha, c, \gamma, m)$ (formulas (4) and (5)) which are given in tables 4 and 5 from data of tables 2 and 3. Actually, the complex gain $F(f)$ (formula (5)) is the characteristic function of a stable probability law of parameters $(\alpha, c, \beta, m')$ (section 3). This strong property is derived from both linearities of ln*att* with $\ln f$ and of $v^{-1}(f)$ with $f^{\alpha-1}$ (or ln$f$), $\alpha$ being the slope of the first line. Values of $\beta$ are given in table

6. They are close to $\pm 1$, and this parameter is closely linked to the causality of filters (section 3-2). Moreover, we show that random propagation times provide the correct framework taking into account the random character of propagation due to irregular motion of molecules.

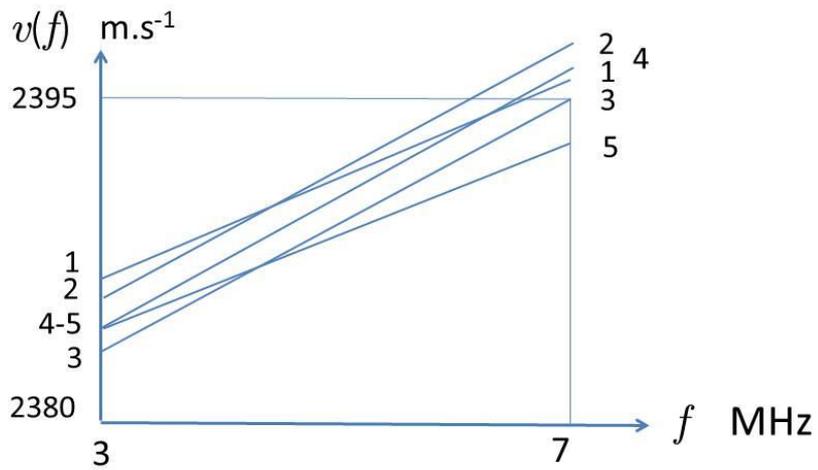

Figure 1a: velocity for bulk samples, linear interpolation

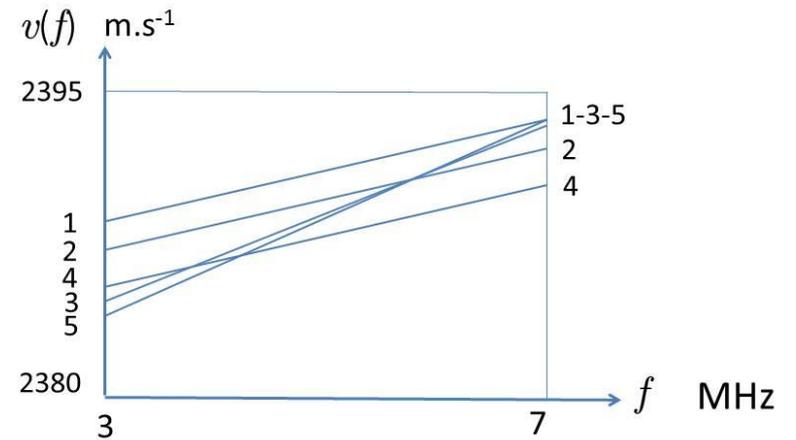

Figure 1b: velocity for surface samples, linear interpolation

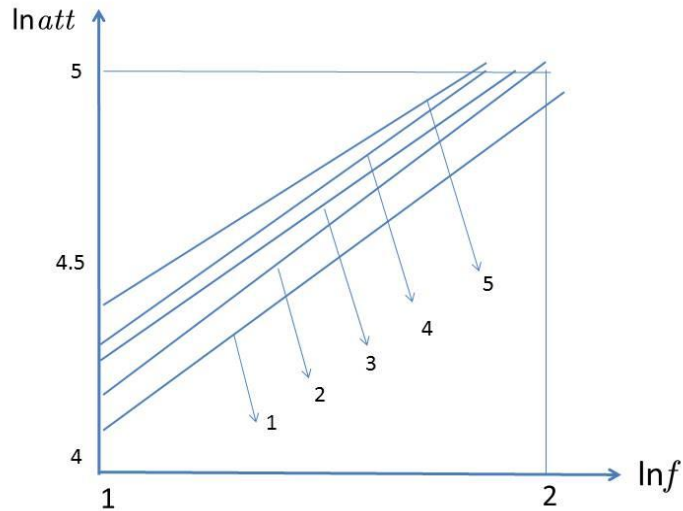

Figure 2a: ln attenuation for bulk samples as function of $\ln f$ ($f$ in MHz)

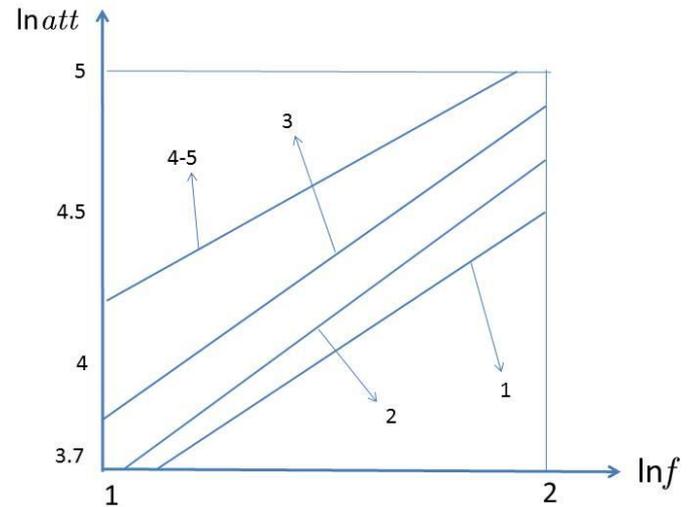

Figure 2b: ln attenuation for surface samples as function of $\ln f$ ($f$ in MHz)

Figures 1a,1b,2a,2b

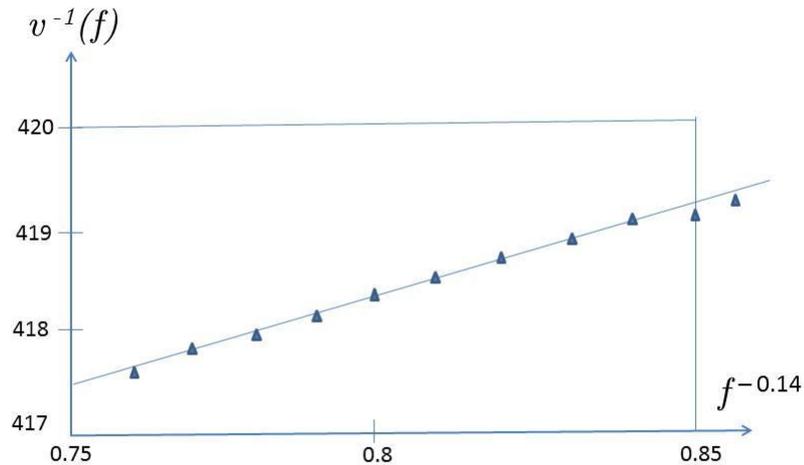

Figure 3a: $v^{-1}(f)$ as a function of $f^{-0.14}$ for bulk sample 1

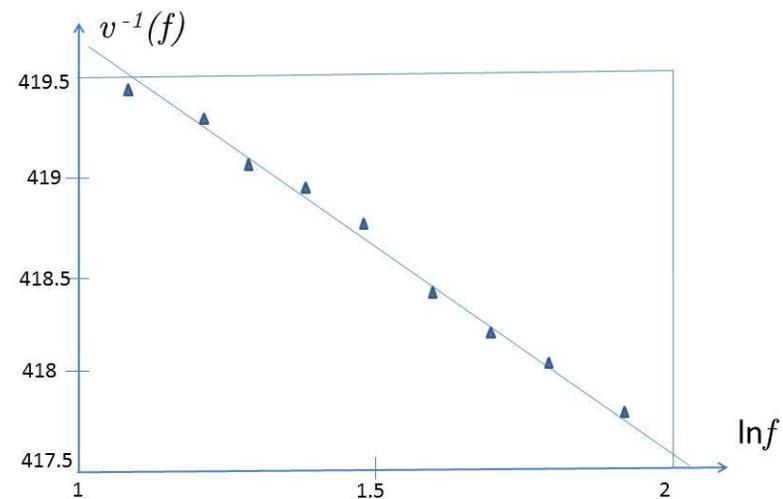

Figure 3b: $v^{-1}(f)$ as a function of $\ln f$ for bulk sample 3

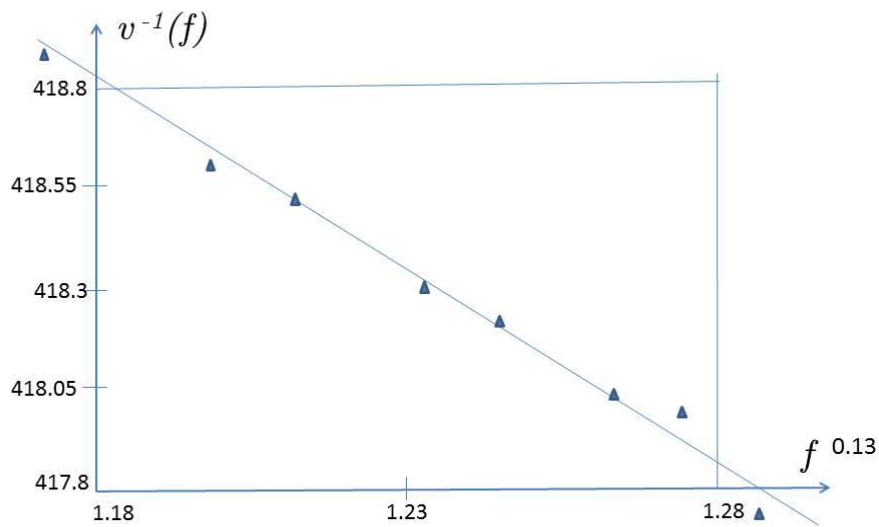

Figure 3c: $v^{-1}(f)$ as a function of $f^{0.13}$ for surface sample 1

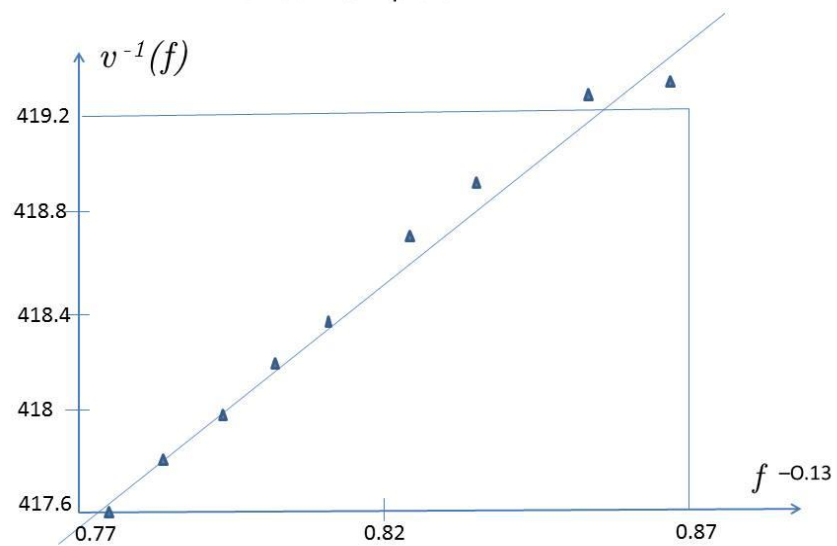

Figure 3d: $v^{-1}(f)$ as a function of $f^{-0.13}$ for surface sample 5

Figures 3a, 3b, 3c, 3d

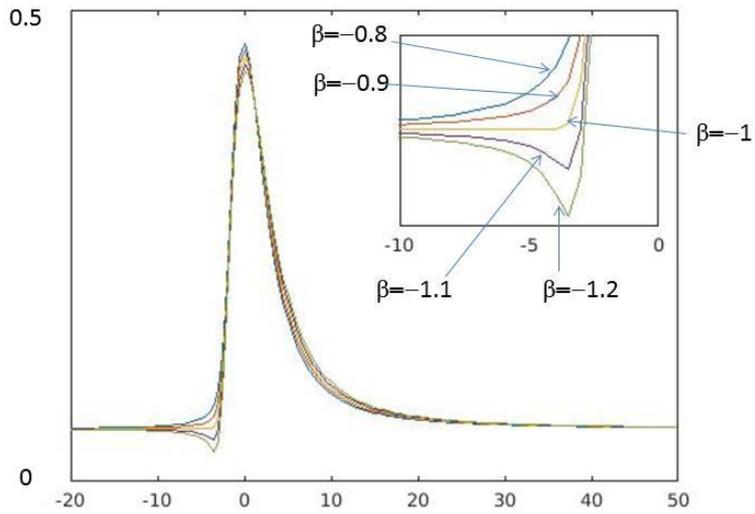

Figure 4a: probability density for α=1, c=2

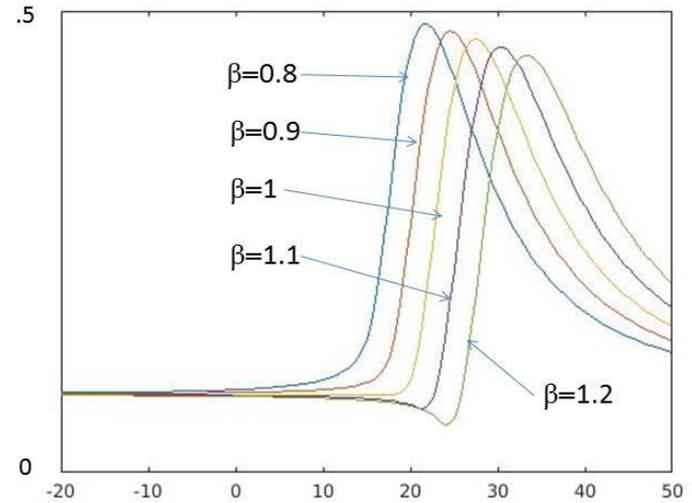

Figure 4b: Probability density for α=0.87, c=5

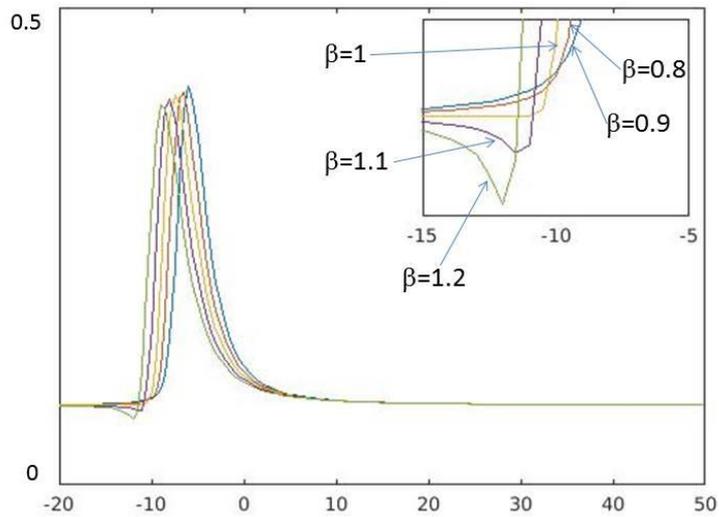

Figure 4c: Probability density for α=1.13, c=1.5

Figures 4a, 4b, 4c